\documentstyle[preprint,aps]{revtex}
\begin{document}
\draft
\title{
Transition radiation
\\
of the neutrino magnetic moment
}
\author{
M. Sakuda\cite{Sakuda}  and Y. Kurihara
}
\address{
National Laboratory for High Energy Physics, KEK,
  Ibaraki-ken 305, Japan
}
\date{\today}
\maketitle
\begin{abstract}
	If the neutrino has a finite mass and a magnetic moment it
would produce transition radiation when
crossing the interface between two media of which plasma frequencies
are $\omega_1$ and $\omega_2$ ($\omega_1 \gg \omega_2$).
We found that the
probability of transition radiation is larger by an order of
magnitude using the quantum theory than that recently reported
by one of us using classical electrodynamics, and
that the energy spectrum of the radiation is uniform up to
 $\sim\gamma \omega_1$,
 where $\gamma$ is the Lorentz
factor of the neutrino ($\gamma=E_{\nu}/m_{\nu}$).
\\
\end{abstract}
\narrowtext

\indent
 	In the standard model \cite{r1} with the right-handed
neutrino singlet ($\nu_R$) the magnetic moment
of the neutrino is induced by radiative corrections, and is estimated
to be negligibly small: $\mu_{\nu}=(3\times 10^{-19}m_{\nu})\mu_{B}$ \cite{r2},
 where $m_{\nu}$ is the neutrino mass in units of eV and
$\mu_{B}$ is the electron Bohr magneton. Thus, the existence of a
neutrino magnetic moment at an order of
$10^{-10}\mu_{B}$ would require a modification of the standard model of the
 electro-weak interaction \cite{r3}. It might also explain the
 solar-neutrino problem \cite{r4,r5,r6}; further, the plasmon decay into
neutrino-antineutrino pair ($\gamma^{\ast}\rightarrow\nu \bar{\nu}$)
would play a more important role in the stellar cooling process \cite{r7}.
The present experimental
upper bounds on the neutrino magnetic moment are
$\mu(\nu_{e})\lesssim 10^{-10}\mu_{B}$ \cite{r8,r9},
$\mu(\nu_{\mu})\lesssim 10^{-9}\mu_{B}$ \cite{r9,r10}, and
$\mu(\nu_{\tau})\lesssim 10^{-6}\mu_{B}$ \cite{r11} at the 90$\%$ CL.
These experimental searches have been performed using
the process of neutrino-electron elastic scattering
 \ \cite{r12} and the  $e^+e^-\rightarrow \gamma \nu \bar{\nu}$ process.
 However, there are other important processes of the
electromagnetic interaction of the neutrino with matter:
Cherenkov radiation and transition radiation. The possibility of
Cherenkov radiation of the neutrino magnetic moment in 1 $km^3$
 of water has recently been studied by Grimus et
al.\ \cite{r13,r14}. The transition radiation of the neutrinos
having a magnetic moment and a
mass was recently discussed by one of us using classical
electrodynamics \cite{r15}.
However, the previous calculation concerning the transition radiation
is not appropriate for the case of neutrinos, since
 such quantum-mechanical effects as the change in the spin
orientation and the recoil of the neutrino during the interaction were
 not taken into account.
In this Letter we revise the calculation of the transition radiation
of a neutrino magnetic moment using quantum theory.
\\
\indent
	Transition radiation (TR) is produced when a charged
particle or a particle with a magnetic moment traverses the
interface between two different media \cite{r16,r17}.
 In quantum theory, the electromagnetic
interaction of the neutrino is described in terms of the Lagrangian density,

\begin{equation}
{\cal L} ={\mu_{\nu} \over 2} \overline{\psi}
\sigma_{\mu \nu}\psi F^{\mu \nu}\ , \label{E1}
\end{equation}
\noindent
where $\mu_{\nu}$ is the magnetic moment defined at the rest
frame of the neutrino, $\psi$ is the
neutrino wave function,
$\sigma_{\mu \nu}={\scriptstyle i \over \scriptstyle 2}
(\gamma_{\mu}\gamma_{\nu}
-\gamma_{\nu}\gamma_{\mu})$, and $F^{\mu \nu}= \partial^{\mu}A^{\nu}-
\partial^{\nu}A^{\mu}$ is the electromagnetic tensor. The
phenomenological quantum theory of the TR of a charged particle
was first given by Garibyan \cite{r18}. It is quite different
from the explanation given by classical
electrodynamics. We will present a calculation of the TR of
the neutrino magnetic moment following Ref.\ \cite{r18}\ \cite{Bell}.
The process is illustrated in Fig.\ \ref{fig1}.
The four-momentum vector of a photon in a
medium having a refractive index of $n$  and satisfying the Maxwell
equations is given by

\begin{equation}
k^{\mu} =(\omega, {\bf k}) \text{, with\ \ }|{\bf k}|=n\omega \ ,\label{E2}
\end{equation}
\noindent
where $\omega$ is the energy of the photon. The magnetic permeability
is assumed to be unity. The effective mass-squared of the photon is
thus given by

\begin{equation}
k^2=(1-n^2)\omega^2 \ . 	\label{E3}
\end{equation}
\indent
	In a uniform medium, the radiation process, $\nu (p_1)\rightarrow
\nu (p_2)+\gamma (k)$, is kinematically allowed at the first order
when $n$ is greater than 1 and $n\beta >1$ is satisfied, where $\beta$ is the
velocity of the neutrino \cite{r19}. This case leads to Cherenkov
radiation of the neutrino magnetic moment. A detailed discussion
for this case can be found elsewhere \cite{r14}.
When the medium is uniform and $n$ is less than 1, the effective
mass-squared of the photon is positive and the radiation process
$\nu \rightarrow \nu +\gamma $ is kinematically forbidden.
However, as can be seen in the following, radiation becomes possible
if there is a plane interface at $z=0$, where the refractive index
suddenly changes from $n_1\ (z<0)$ to $n_2\ (z>0)$.
A transition probability for the radiation process
$\nu \rightarrow \nu +\gamma $ at the lowest order
is calculated by using formula \cite{r20},
\begin{equation}
\Gamma = \bigl|\ S_{fi}\ \bigr|^2 {Vd^3{\bf p}_2 \over (2\pi)^3}
{Vd^3{\bf k} \over (2\pi)^3}\ ,\label{E4}
\end{equation}
\begin{equation}
\text{	with } S =i \int d^4x\ {\cal L}\ ,\label{E5}
\end{equation}
where $S$  is the $S$ matrix, $V =L^3$ is the spacial volume of the
interaction region and ${\cal L}$ is the Lagrangian given
in Eq.\ (\ref{E1}). We assume that the wave funtions
describing the initial-state
and final-state neutrino are given by

\begin{equation}
\psi_i (x)=\sqrt{ {m_{\nu}\over E_i V} } u(p_i,\lambda_i)\exp
(-ip_i\cdot x)\ ,\ (i=1,2) \ , \label{EX}
\end{equation}
\noindent
where $m_{\nu}$ is the neutrino mass,  $E_i $ is the
neutrino energy,  and $u(p_i,\lambda_i) $ denotes a
positive-energy solution of the Dirac equation with
four-momentum $p_i^{\mu}$ and
helicity $\lambda_i$.
Each of the wave functions $\psi_i (x)\  (i=1, 2) $ is normalized
to unit probability in a box of volume $V$.
The $S$ matrix is calculated from Eqs.\ (\ref{E1}) and (\ref{E5}) as

$$\bigl|\ S_{fi}\ \bigr|^2=(2\pi)^3L^2T \delta (p_{1x}-p_{2x}-k_x)
\delta (p_{1y}-p_{2y}-k_y) \delta (E_1-E_2-\omega) $$

\begin{equation}
\cdot {m_{\nu}\over E_1V}{m_{\nu}\over E_2V} {1\over 2\omega n^2V}
\Bigl| \int_{-L/2}^{L/2}\ dz\ \exp [i(p_{1z}-p_{2z}-k_z)z ]\ M_{fi}\
\Bigr|^2\ ,\label{E6}
\end{equation}
\begin{equation}
{\rm with \ \ } M_{fi} ={\mu_{\nu} \over 2}\bar u(p_2,\lambda_2)
\sigma_{\mu \nu} u(p_1,\lambda_1) i(k^{\mu}\varepsilon^{\nu}-
k^{\nu}\varepsilon^{\mu}) , \label{E7}
\end{equation}
\noindent
where  $\varepsilon^{\mu}$ is the unit polarization vector
of the photon satisfying $k\cdot\varepsilon=0$ and
$T$ is the time interval of the observation
($L=\beta T$). In connection with the phase in the integrand of
Eq.\ (\ref{E6}), the formation-zone length of the medium is defined as

\begin{equation}
Z(n)\equiv (p_{1z}-p_{2z}-k_z)^{-1} =(p_{1z}-p_{2z}-n\omega\cos
\theta )^{-1}\ , \label{E8}
\end{equation}

\noindent
where $\theta$ is the angle between the photon and the direction of the
incident neutrino. The integral of Eq.\ (\ref{E6}) must be performed for
the ($-L/2,0$) and ($0,L/2$) regions separately.
Since the integrand oscillates beyond the depth of the
formation-zone length ($z \ll -Z(n_1)$
or $z \gg Z(n_2)$), the contribution of the lower and
upper limits ($z =\pm L/2$) of the integral can be neglected
($L\gg Z(n_i)$ is assumed). Only radiation from the volume near
to the interface ($-Z(n_1)\lesssim z\lesssim Z(n_2)$ ) is
added coherently. This is the case with TR.
 A fraction of the momentum (z-component) of the neutrino is lost
in the volume near to the boundary between the two media.
A detailed discussion of the energy-momentum (non-)
conservation in the process of TR can be found in Ref.\cite{r17}.
We obtain the energy intensity $S$
per interface from Eqs.\ (\ref{E4}) and (\ref{E6}) as

\begin{equation}
{d^2S \over d\theta d\omega}\equiv \omega{d^2\Gamma \over d\theta d\omega}
 ={\mu_{\nu}^2\omega ^2\sin\theta \over
8\pi^2\beta\beta_2\gamma\gamma_2 }\bigl| A_1\ -\ A_2\ \bigr|^2\ ,
\label{E9}
\end{equation}
\begin{equation}
{\rm with\  } A_{\alpha}\equiv {1 \over n_{\alpha}}\bar u(p_2,\lambda_2)
{\not \! k_{\alpha} \not \! \varepsilon
\over p-p_{2z}-n_{\alpha}\omega\cos\theta }
 u(p_1,\lambda_1)\ ,\ \  (\alpha=1,2)\ . \label{E10}
\end{equation}
\noindent
 If the momentum of the incident
neutrino, $p_1^{\mu}=(E_{\nu},0,0,p)$, is given,
the other quantities in Eqs.(\ref{E9}-\ref{E10})
are calculated from the following equations:

\begin{equation}
E_2=E_{\nu}-\omega\ ,
p_{2z}=\sqrt{E_2^2 - m_{\nu}^2 -  n_{\alpha}^2\omega^2 \sin ^2\theta }\ ,
\beta_2=p_{2z}/E_2\ ,
{\rm and\ \ } \gamma_2=E_2/m_{\nu}\ .  \label{E11}
\end{equation}

\noindent
Since we are interested in the radiation in the x-ray region
 ($ n_{\alpha}(\omega)\sim 1$),
we assume that the refractive index can be expressed
in terms of the plasma frequencies $\omega_{\alpha}\ (\alpha=1,2)$
as $ n_{\alpha}(\omega)=1-\omega_{\alpha}^2/2\omega^2$ for
$\omega \gg \omega_{\alpha}$, and that the radiation
from medium 1 ($z<0$) propagates through the interface without
any reflection or refraction. Thus, variables $\theta$ and $\varepsilon^{\mu}$
 are independent of the medium, $\alpha$.

\indent
	We show the energy spectrum and the total energy per interface in
Figs.\ \ref{fig2}-\ref{fig3}
for the typical parameters: $E_{\nu}=1\ MeV$, $\omega_1=\omega_p=20\
eV$ (polypropylene) and  $\omega_2=0\ eV$ (vacuum).
In the calculation we average Eq.\ (\ref{E9}) over
the helicity states of the incident neutrino,
sum it over the helicity states of the outgoing neutrino and
sum it over two polarization states of the radiated photon.
The probability is found to be the same as that in which the
incident neutrino has a definite helicity of $\lambda_1$=-1 or 1. The
calculations of Eqs.(\ref{E9}-\ref{E10}) are performed numerically using
the helicity amplitude subroutines \cite{r21}\ \cite{r22}.
The total energy $S$  is obtained by integrating Eq.\ (\ref{E9})
over the $\omega$ and $\theta$
 ranges, (0, $E_{\nu}-m_{\nu}$) and (0, $\pi/2$), respectively \cite{r23}.
 A previous calculation using classical theory \cite{r15} is also
shown for a comparison. The features of the TR of a neutrino
magnetic moment are summarized as follows:(a) the majority of the
radiation comes from the helicity-flip amplitude,
 and, thus, the effect is purely quantum mechanical; (b) the energy spectrum
is flat up to 0.5$\gamma \omega_p$, and then decreases rapidly;(c) the energy
 intensity is proportional to the Lorentz factor ($\gamma$)
for $\gamma \omega_p \ll E_{\nu}$ (i.e. $m_{\nu} \gg \omega_p$ )
and begins to saturate for $\gamma \omega_p > E_{\nu}$ (i.e.
 $m_{\nu} < \omega_p$ ):
\begin{mathletters}
\label{E12:all}
\begin{equation}
S=1.7\times 10^{-12}(\mu_{\nu}/\mu_B)^2\gamma \omega_p \ \
{\rm  for\ }
\gamma \omega_p \ll E_{\nu}\ ,  \label{E12:a}
\end{equation}
\begin{equation}
\ \  =4.5\times 10^{-13}(\mu_{\nu}/\mu_B)^2E_{\nu}\ \   {\rm  for\ }
\gamma \omega_p \gg E_{\nu}\ .	 \label{E12:b}
\end{equation}
\end{mathletters}

\noindent
The coefficients (=probability) originate from a dimension-less constant
$\mu_B^2\omega_p^2=$ $3.5\times 10^{-11}$ for $\omega_p=20$ eV \
\cite{r24}; and (d) the emitted angle has a peak
at the forward direction, $ \theta \sim 1/\gamma$.

\indent
First of all, the energy intensity turns out to be larger by an order
of magnitude than that
 ($S=1.9 \times 10^{-13}(\mu_{\nu}/\mu_B)^2\gamma \omega_p $ for
$\gamma \omega_p \ll E_{\nu}$)
estimated by classical theory. The TR yield is not reduced even
for the case of a small mass under the condition that the
magnitude of the magnetic moment is the same. The recoil effect
becomes important for $\gamma \omega_p \gtrsim E_{\nu}$.
A dominant helicity-flip
amplitude is characteristic of the interaction of
Eq.\ (\ref{E1}), which has already been pointed out concerning other processes
\cite{r12,r13}. A previous calculation using classical theory
corresponds to the helicity-nonflip transition.
To confirm this point, we also show the energy spectrum and
the total intensity in Figs.\ \ref{fig2}-\ref{fig3} using quantum theory for
the case when the helicity is not changed during the radiation
process, i.e. $\lambda_1=\lambda_2=$-1 (dashed-dot line).
 The calculation using quantum theory takes into account the recoil
effect, i.e. $p_2\not =p_1$.
The present calculation for the process disagrees
with that of classical theory only for the region
($\omega \sim E_{\nu}$) where the recoil effect
is important.
The classical calculation corresponds to
the radiation of a particle with such a large magnetic moment (or spin)
and large mass that the radiation has
no effect on the spin state or the trajectory of the particle.
\\
\indent
 The sensitivity of a typical transition radiation detector has already
been discussed \cite{r15}. The present work shows that the
TR yield has increased by about 10, and that the
sensitivity of the method to the neutrino magnetic moment
for a small mass region ($m_{\nu}<\omega_p$) is not as much decreased as
that given previously. We now present a calculation
of the TR yield for a practical detector containing many foils, where
the interference effects between the individual interfaces
(=``formation-zone effect'') must be taken into account  \cite{r25}.
For example, the TR yield per interface
given in Eq.\ (\ref{E9})
must be corrected for a periodic radiator comprising $N$
polypropylene foils ($N$=100$\sim$500, $\omega_1$=20 eV and thickness
$\ell_1$=0.1 mm) stretched in air ($\omega_2$=0.8 eV and spacing
$\ell_2$=2 mm)\ \cite{r26}. The average TR yield per
interface at $E_{\nu}=$1 MeV is estimated to be
almost the same for $m_{\nu}\gg \omega_1$
 and  about a half for $0.01\ eV<m_{\nu}<\omega_1$
 as compared to that given in Eq.\ (\ref{E9}).
 The reduction due to the formation-zone effect is not so large in this
case\ \cite{r28}.
\\
\indent
	In conclusion, we have revised the calculation of the transition
radiation of a neutrino magnetic moment using quantum theory,
 where both the helicity-flip effect  and the recoil effect
 are taken into account. We found that it is larger by an order
of magnitude than that estimated using classical electrodynamics
and that the energy spectrum of the radiation is uniform up to
 $0.5\gamma \omega_p$. The transition radiation of the
neutrino magnetic moment is unique in that the energy
intensity depends explicitly on the neutrino mass.
\\
\indent
 This work would not have been completed without the valuable comments of
 Makoto Kobayashi (KEK). It is a pleasure to acknowledge  K. Hagiwara
and  Y. Shimizu (KEK) for both encouragement and useful discussions.
 We would also like to
thank T. Tsuboyama, R. Enomoto, S. Kawabata, A. Miyamoto,
J. Fujimoto, and J. Kanzaki (KEK) for
useful advice during the work.
\\

\protect
\begin{figure}
\caption{
 Transition radiation at the interface of two
media: $\nu (p_1)\rightarrow \nu (p_2)+\gamma(k)$. The refractive
index changes from $n_1$  to $n_2$ at $z=0$.
 }
\label{fig1}
\end{figure}

\protect
\begin{figure}
\caption{
 Energy spectrum of the TR of the neutrino magnetic
moment ($\mu_{\nu}=\mu_B$) for $E_{\nu}$=1 MeV,
$m_{\nu}$=100 eV (solid line), and $m_{\nu}$=0.1 eV (dashed
line). The plasma frequencies of media 1 and 2 are $\omega_1$=20 eV and
$\omega_2$=0 eV, respectively. A calculation using classical electrodynamics
\protect\cite{r15} is also
shown by the dotted line ($m_{\nu}$=100 eV), while the dashed-dot line
indicates that of quantum theory when the helicity is not changed
 during the interaction, but the recoil effect is taken into account. }
\label{fig2}
\end{figure}

\begin{figure}
\caption{
 Total TR energy of the neutrino magnetic
moment ($\mu_{\nu}=\mu_B$) as a function of the mass for
$E_{\nu}$=1 MeV (solid line). That using classical electrodynamics
\protect\cite{r15} is indicated by the dotted line,
 while the dashed-dot line is that
of quantum theory when the helicity is not changed during the
interaction, but the recoil effect is taken into account. }
\label{fig3}
\end{figure}


\begin{references}
\bibitem{r14} Electromagnetic interaction of the neutrino in the
Kamiokande detector has been studied in Ref.\ \cite{r6}
in the more dominant $\nu e\rightarrow\nu e$ process
 with the recoil electron producing Cherenkov radiation.
\bibitem{r19} We use the Heaviside-Lorentz units and set $c=\hbar =1$
 throughout the paper.
\bibitem{r22} For the region $m_{\nu}\gg \omega_p$, the analytic form of
Eq.\ (\ref{E9}) is given by

$${ d^2S \over d\theta d\omega} ={\mu_{\nu}^2\omega_p^4
\sin\theta\cos\theta (1-\beta\beta_2\cos^2\theta+{1\over
\gamma\gamma_2}) \over 16\pi^2\beta\beta_2(p-p_{2z}-n\omega\cos
\theta)^2
(p-p_{2z}-\omega\cos \theta) }\  .$$
\bibitem{r24} The coefficients of Eq.\ (\ref{E12:all}) depend slightly on
the incident energy.
 For example, the numbers in Eqs.(\ref{E12:a}) and (\ref{E12:b})
at $E_{\nu}$=1 GeV are $1.8\times 10^{-12}$ and $6.8\times 10^{-13}$,
respectively.
\\
\bibitem{r26}  Total TR yield $S(N)$ from N foils with thickness
$\ell_1$ and spacing $\ell_2$ is given by \ \cite{r25}

$${ d^2S(N) \over d\theta d\omega} ={ d^2S \over d\theta d\omega}
\cdot 4 \sin^2({\ell_1 \over 2Z_1})\cdot { \sin^2[N({\ell_1 \over 2Z_1}+
{\ell_2 \over 2Z_2})] \over \sin^2({\ell_1 \over 2Z_1}+{\ell_2 \over
2Z_2}) }\ ,\ $$
where $Z_1$ and  $Z_2$ are the formation-zone length of the two media
and  $S$ is the TR yield per interface given in Eq.\ (\ref{E9}).
 The average TR yield  per interface is given by $S(N)/(2N)$.
\bibitem{r28}  We note that the gap length is usually smaller
than the formation-zone length of the gaps for $\gamma \gg 10^5$
in most transition-radiation detectors and
that the main formation-zone effect comes from the gaps and
not from the foils.
\end{references}
\end{document}